\documentclass[aps,pra,twocolumn,amsmath,showpacs,amssymb,floatfix,nofootinbib,superscriptaddress,groupedaddress]{revtex4}
\usepackage{graphicx,psfrag,bm,times}
\begin{document}
\title{Separability in Asymmetric Phase-Covariant Cloning}
\author{A. T. Rezakhani }
\affiliation{Department of
Physics, Sharif University of Technology, P. O. Box 11365-9161,
Tehran, Iran}
\author{S. Siadatnejad}
 \affiliation{Department of
Mathematical Sciences, Sharif University of Technology, P. O. Box
11365-9415, Tehran, Iran}
\author{A. H. Ghaderi}
 \affiliation{Department of Physics, Amirkabir University of Technology, P. O. Box 15914, Tehran,
Iran}
\begin{abstract}
Here, asymmetric phase-covariant quantum cloning machines are
defined and trade-off between qualities of their outputs and its
impact on entanglement properties of the outputs are studies. In
addition, optimal families among these cloners are introduced and
 also their entanglement properties are investigated. An explicit proof
of optimality is presented for the case of qubits, which is based
on the no-signaling condition. Our optimality proof can also be
used to derive an upper bound on trade-off relations for a more
general class of optimal cloners which clone states on a specific
orbit of the Bloch sphere. It is shown that the optimal cloners of
the equatorial states, as in the case of symmetric phase-covariant
cloning, give rise to two separable clones, and in this sense
these states are unique. For these cloners it is shown that total
output is of \textsc{ghz}-type.
\end{abstract}
\date{\today}
\pacs{03.67.-a, 03.67.Mn, 89.70.+c} \maketitle
\section{Introduction}{\label{sec1}}
In quantum information theory the so-called {\em no-cloning
theorem} prohibits ideal copying of a quantum state
{\cite{Dieks,WZ,NoBroadcast}}.  Although it may seem a restrictive
fact, it is an advantage over classical information theory.
Because security in quantum cryptography is largely attributed to
impossibility of exact copying of quantum data. It is impossible
to clone quantum information exactly, however, it is useful to
know how well one can achieve this goal. As well, there are
inevitable needs to investigate this since, for example, storage
and retrieval of quantum information on quantum computers are
essentially related to copying . Hence quantum cloning enters into
the scope of real experiments. Also, as is seen any success in
good cloning makes quantum cryptography more at risk. For example
in BB84 protocol {\cite{Bennett}} there is a link between optimal
cloning of equatorial states and optimal eavesdropping attack.
There have been extensive studies on this subject that illuminate
some aspects of quantum cloning
{\cite{HB1,HB2,Gisin,Bruss1,Werner1,Werner2,Bruss2,Zanardi,Dlevel}}.

In this paper, we investigate a family of asymmetric cloning
machines for $d$-dimensional states in the form of
$|\psi\rangle=\frac{1}{\sqrt{d}}\sum_ke^{i\phi_k}$. However
asymmetric machines are important in their own merits, their
occurrence in the context of quantum information theory can also
be attributed to the situations in which one of clones needs to be
a bit better than the other, or when there may be an internal flaw
in hardware of symmetric cloner that makes two copies
non-identical. Also in studying these machines various no-cloning
inequalities, which are consequences of the quantum uncertainty
principle, are relevant and obtain practical meanings. We also
investigate these machines in the sense of entanglement produced
in their outputs.

The paper is structured as follows.  In Sec. \ref{sec2} we review
 universal asymmetric cloning machines, based on trivial
asymmetrization of the original universal cloning machine of
Bu\v{z}ek and Hillery {\cite{HB1}}.  Indeed, this is nothing more
than a simple re-explanation of the asymmetric cloners firstly
introduced by Cerf {\cite{Cerf0,Cerf1,Cerf2}}. Its general
properties are reviewed and a trade-off relation for the qualities
of the two clones is derived. In Sec. \ref{sec3}, after some
general remarks on $d$-dimensional asymmetric phase-covariant
cloners, we study a special purpose asymmetric machine for cloning
$x-y$ equatorial states of the Bloch sphere; phase-covariant qubit
cloners. Next, we investigate optimality of such asymmetric
machines. Then we analyze the separability of the output copies
and show that among all inputs only equatorial states give rise to
two separable outputs, that is, output clones of an optimal
phase-covariant cloner are unentangled.  In addition, it is shown
that total outputs of these machines are of \textsc{ghz}-type. The
paper is concluded in Sec. \ref{sec4}.

\section{universal asymmetric cloning machine}{\label{sec2}}

In this section we are going to devise an asymmetric cloning
machine which is universal, that is, it treats all inputs in the
same way.  For the sake of simplicity, here, we restrict ourselves
to the case of duplicators, i.e. 1$\rightarrow$2 universal
cloners. However, extension to triplicators is also
straightforward. The question of how well one can design an
approximate duplicator of a qubit (or qudit), provided that the
qualities of the two outputs be independent of the input states,
has been investigated by Bu\v{z}ek and Hillery {\cite{HB1,HB2}}
and the others
{\cite{Gisin,Bruss1,Werner1,Werner2,Bruss2,Zanardi,Dlevel}}.

\begin{figure}
\psfrag{eta1}[Bc][][0.75][0]{$\eta_{A}$}
\psfrag{eta2}[Bc][][0.75][0]{$\eta_{B}$}
\includegraphics[width=4.7cm,height=4.6cm]{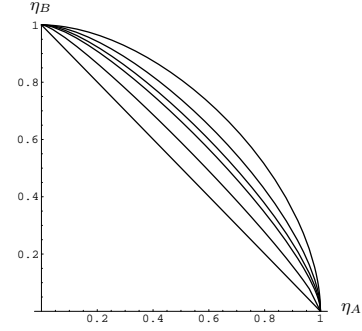}
\caption{Trade-off ellipses. Each graph corresponds to a specific
dimension, for example the straight line is for infinite
dimensional case. } \label{fig1}
\end{figure}

At first, we briefly review $d$-dimensional universal cloning
machines following Bu\v{z}ek and Hillery {\cite{HB2}}. Consider
the unitary transformation
\begin{eqnarray}
&|i\rangle_{A}|0\rangle_{B}|\Sigma\rangle_{X} \longrightarrow
\mu|i\rangle_{A}|i\rangle_{B}|i\rangle_{X}\nonumber\\
&+\nu\sum_{j\neq
i}^{d-1}(|i\rangle_{A}|j\rangle_{B}+|j\rangle_{A}|i\rangle_{B})~|j\rangle_{X},
 \label{eq1}
\end{eqnarray}
in which {\small{$A$}} and {\small{$B$}} are, respectively, input
and blank qudits, and {\small{$X$}} is an ancilla that always can
be considered as the cloning machine itself which is initially in
a fixed state, say $|\Sigma\rangle$.  The set $ \{|i\rangle_{A(X)}
\} _{i=0}^{d-1}$ is a set of orthonormal basis vectors of the
Hilbert space of input (machine); $\mathcal{H} _{A(X)}$. Without
loss of generality, $\mu$ and $\nu$ can always be considered to be
real parameters. Requiring unitarity of the transformation and the
following conditions: (i) quality of cloning (defined based on
fidelity of the copies $F:=\langle\psi|\rho^{(out)}|\psi\rangle$)
does not depend on the particular state which is going to be
copied ({\em{universality}} or input state-independence), (ii) the
outputs are {\em{symmetric}}, i.e.
$\rho_{A}^{(out)}=\rho_{B}^{(out)}$, the following relations can
be obtained
\begin{subequations}
\begin{eqnarray}
   &\rho_{A(B)}^{(out)}=\eta\rho_{A(B)}^{(id)}+\frac{1-\eta}{d}1_{A(B)}, \label{eq2a}\\
   &\mu^2=2\mu\nu,\mu^2=\frac{2}{d+1},\nu^2=\frac{1}{2(d+1)} \label{eq2b}\\
   &\eta=\mu^2+(d-2)\nu^2=\frac{d+2}{2(d+1)},\label{eq2c}
\end{eqnarray}
\end{subequations}
where $1_{A(B)}$ stands for the $d\times d$ identity operator on
the space of ${\cal{H}}_{A(B)}$, and $\eta=\frac{dF-1}{d-1}$ is
called shrinking factor. Some points on the above cloning
transformation are worth noting. In $d=2$, this machine simply
reduces to the original universal cloning machine \cite{HB1}, with
$\eta=\frac{2}{3}$ or $F=\frac{5}{6}$. This machine is proved to
be optimal in that it produces maximal fidelity considering its
requirements \cite{Bruss2,Werner1,Werner2,Zanardi,Dlevel}.  It can
be justified that symmetry of the outputs is a consequence of
equality of the coefficients of the terms
$|ij\rangle_{AB}|j\rangle_{X}$ and $|ji\rangle_{AB}|j\rangle_{X}$
in Eq.~(\ref{eq1}). Thus one can consider it as a starting point
for extension to transformations which produce asymmetric output
copies. Here, it must be noted that this kind of survey is not
something different from the asymmetric cloning introduced by Cerf
\cite{Cerf0, Cerf1,Cerf2}. However, to clarify the subject we use
a simpler exposition.  Let us start simply by giving different
contributions to the two latter terms. Hence, the following
cloning transformation can be introduced (which is an isometry)
\begin{eqnarray}
&|i\rangle_{A}|0\rangle_{B}|\Sigma\rangle_{X} \longrightarrow
\mu|i\rangle_{A}|i\rangle_{B}|i\rangle_{X}\nonumber\\
&+\nu\sum_{j\neq
i}^{d-1}|i\rangle_{A}|j\rangle_{B}|j\rangle_{X}+\xi\sum_{j\neq
i}^{d-1}|j\rangle_{A}|i\rangle_{B}|j\rangle_{X}. \label{eq3}
\end{eqnarray}
If a state in the form of
$|\psi\rangle=\sum_{i}\alpha_{i}|i\rangle$ is given to the machine
as an input, then, the state of the output copy {\small{$A$}}
becomes
\begin{eqnarray}
&\rho_{A}^{(out)}=|\psi\rangle_{A}\langle\psi|[(d-2)\nu^2+2\mu\nu]+\xi^{2}1_{A}\nonumber\\
&+
(\mu^2+\nu^2-\xi^2-2\mu\nu)\sum_{i}|\alpha_{i}|^2|i\rangle_{A}\langle
i|, \label{eq4}
\end{eqnarray}
and similarly for the copy {\small{$B$}} (with the replacements
$\xi\leftrightarrow\nu$,
{\small{$A$}}$\leftrightarrow${\small{$B$}}).  The last term in
Eq.~(\ref{eq4}) is obviously state-dependent.  If we impose
state-independence condition for the cloning machine, by
considering the fact that necessary and sufficient conditions for
$F_{A}$ being state-independent is that for an input state
$\rho^{(id)}$, the output state $\rho_{A}^{(out)}$ has a form as
\begin{eqnarray}
&\rho_{A}^{(out)}=\eta_{A}\rho_{A}^{(id)}+\frac{1-\eta_{A}}{d}1_{A},
\label{eq5}
\end{eqnarray}
we get the following relations
\begin{subequations}
\begin{eqnarray}
&\eta_{A}=(d-2)\nu^2+2\mu\nu \label{eq6a}\\
&\mu^2+(d-1)(\nu^2+\xi^2)=1
\label{eq6b}\\
&\frac{1-\eta_{A}}{d}=\xi^2\label{eq6c}\\
&\mu^2+\nu^2-\xi^2-2\mu\nu=0. \label{eq6d}
\end{eqnarray}
\end{subequations}
Eqs.~(\ref{eq6b}) and (\ref{eq6d}), and the corresponding relation
for the copy {\small{$B$}}, $\mu^2+\xi^2-\nu^2-2\mu\xi=0$, give
rise to the following result
\begin{eqnarray}
 &  \mu=\nu+\xi. \label{eq7}
\end{eqnarray}
If we introduce the parameterization: $\nu=r\cos\phi$ and
$\xi=r\sin\phi$, the expressions for $F_{A}$ and $F_{B}$ take the
simple forms as below
\begin{subequations}
\begin{eqnarray}
   &F_{A}=\frac{d\cos^2\phi+\sin^2\phi+\sin 2\phi}{d+\sin 2\phi}\label{eq8a}\\
   &F_{B}=\frac{d\sin^2\phi+\cos^2\phi+\sin 2\phi}{d+\sin 2\phi},\label{eq8b}
\end{eqnarray}
\end{subequations}
from which by cancelation of $\phi$ we reach a relation between
the fidelities as below
\begin{eqnarray}
&F_{A}^{2}+F_{B}^2+2\frac{d^2-2}{d^2}F_{A}F_{B}-2\frac{d^2+d-2}{d^2}(F_{A}+F_{B})\nonumber\\
&+\frac{(d-1)(d+3)}{d^2}=0.\label{eq9}
\end{eqnarray}
This is equation of a set of ellipses (in the space of fidelities)
that their eccentricities vary with dimension.  Using the relation
$\eta=\frac{d~F-1}{d-1}$ a corresponding set of ellipses in the
space of shrinking factors can be found.  Figure \ref{fig1}
illustrates these ellipses for some specific dimensions. As is
seen in infinite dimensional case the corresponding ellipse
shrinks to the line $\eta_{A}+\eta_{B}=1$, and also all of
ellipses in $(\eta_{A}=0,\eta_{B}=1)$ point are tangent to
$\eta_{B}=1$ line (and similarly, in $(1,0)$). Also the slope of
the tangents to
the ellipses in the points on the symmetry axis ($\eta_{A}=\eta_{B}$) is 1.\\
In a given dimension, the corresponding ellipse, indeed, induces a
kind of complementarity or trade-off between the two copies, since
if one fixes one of the parameters the other one is determined
too. So we have provided a graph for full spectrum of the
qualities, in which special points corresponding to
$\eta_{A}=\eta_{B}$ on each graph are representatives of the symmetric cloners.\\
Now, we can investigate question of optimality of this simple
cloning transformation.  Similar questions have been studied
earlier in {\cite{HB3,HB4,Cerf0,Cerf1,Cerf2,Niu}}.  As we stressed
above, it can be seen that this machine is nothing but a
Heisenberg cloning machine which is firstly introduced by Cerf
{\cite{Cerf1,Cerf2}}. The quantum uncertainty principle gives rise
to the so-called no-cloning inequalities which are upper bounds on
trade-off relations for the qualities of the two clones.
Heisenberg cloning machines generally give two non-identical
output copies, each of which comes out of a different Heisenberg
channel. It can be simply checked that the cloning transformation
(\ref{eq3}) saturates the no-cloning inequality, and, therefore,
it corresponds to an optimal cloner.

In the next section we relax the universality condition and focus
on asymmetric phase-covariant cloning machines.

\section{Asymmetric Phase-covariant cloning machines }{\label{sec3}}
It is evident that if one {\em a priori} has a partial knowledge
about input states, then utilizing this information, more
efficient special purpose cloning machines can be designed. This
fact leads to the investigation of some state-dependent cloners,
among which the class of phase-covariant cloners lies.  These
special cloners are designed to clone equatorial states as well as
possible (better than the universal cloning).  This class, in the
case of symmetric cloning, has been studied previously
{\cite{Bruss3,D'Ariano,PCCM,DPCCM,Ours}}.  In this section we
investigate asymmetric version of phase-covariant machines.

In Eq.~(\ref{eq4}) it is seen that the last term,
$\sum_{i}|\alpha_{i}|^2|i\rangle\langle i|$, depends on input
state of the machine.  In the special case of input states in the
form
\begin{eqnarray}
&|\psi\rangle=\frac{1}{\sqrt{d}}\sum_{k}e^{i\phi_{k}}|k\rangle,~~~(0\leq
\phi_k < 2\pi)\label{eq14}
\end{eqnarray}
which are covariant with respect to rotations of the phases, this
term {\em automatically} reduces to the identity matrix and,
therefore, becomes state-independent. As a result, for this class
of inputs one does not need to consider the conditio
~(\ref{eq6d}), since this condition was to cancel the contribution
of the state-dependent term. Insisting on having this condition
results in a less efficient cloner for this particular class of
inputs. Anyway, here Eq.~(\ref{eq6d}) is not necessary and
Eq.~(\ref{eq4}) reduces to
\begin{eqnarray}
&\rho^{(out)}_{A}=|\psi\rangle_{A}\langle
\psi|[\nu^2(d-2)+2\mu\nu]\nonumber\\
&\hskip
15mm+(\xi^2+\frac{\mu^2+\nu^2-\xi^2-2\mu\nu}{d})1_{A},\label{eq15}
\end{eqnarray}
and similarly for the {\small{$B$}} copy (by the simple
replacement of $\nu\leftrightarrow \xi$).  Then, shrinking factors
of the two output clones are
\begin{subequations}
\begin{eqnarray}
&\eta_{A}=2\mu\nu+(d-2)\nu^2,\label{eq16a}\\
&\eta_{B}=2\mu\xi+(d-2)\xi^2.\label{eq16b}
\end{eqnarray}
\end{subequations}
Using the normalization condition (\ref{eq6b}), the above
equations can be simplified as below
\begin{subequations}
\begin{eqnarray}
&\eta_{A}=(d-2)\nu^2+2\nu\sqrt{1-(d-1)(\nu^2+\xi^2)},\label{eq17a}\\
&\eta_{B}=(d-2)\xi^2+2\xi\sqrt{1-(d-1)(\nu^2+\xi^2)}.\label{eq17b}
\end{eqnarray}
\end{subequations}
As is seen, contrary to the case of universal cloning, here we are
left with two free tuning parameters to set.
\begin{figure}[tp]
\psfrag{x}[Bc][][0.75][0]{$\eta_{A}$}
\psfrag{y}[Bc][][0.75][0]{$\eta_{B}$}
\includegraphics[width=5cm,height=5cm]{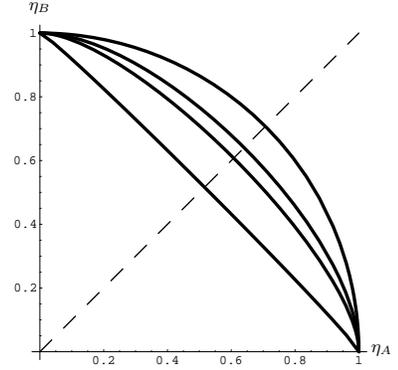}
\caption{Trade-off ellipses in the asymmetric phase-covariant
cloning machine for $d$=2,3,4,30,  respectively from top to down.
} \label{fig2}
\end{figure}

Now, in this class of cloners an {\em optimal} machine is defined
as the following: in this machine if we fix the quality of one of
the clones--say {\small{$A$}}--then the quality of the
other clone is the highest possible value.\\
From Eq.~(\ref{eq17a}) one can find $\xi$ in terms of $\nu$ and
$\eta_A$. After inserting this value in Eq.~(\ref{eq17b}), we have
\begin{eqnarray}
\aligned
\eta_{B}(\nu)&=\frac{d-2}{d-1}[1-(d-1)\nu^2-(\frac{\eta_{A}-(d-2)\nu^2}{2\nu})^2]\\
&\quad+\frac{\eta_A-(d-2)\nu^2}{\nu}\sqrt{\frac{1-(\frac{\eta_{A}-(d-2)\nu^2}{2\nu})}{d-1}-\nu^2}.\label{eq18}
\endaligned
\end{eqnarray}
By optimization of $\eta_{B}(\nu)$ (and assuming
$\eta_{A}$=const.) the value of $\nu_{optimal}(\eta_{A})$ can be
found for any dimension $d$. Unfortunately, in general, there is
not a closed analytical form for this value. However, the correct
solutions can be found by a simple numerical examination.
Figure~\ref{fig2} shows the trade-off diagrams ($\eta_{A}$ vs.
$\eta_{B}$) for some typical dimensions. It can be inferred that
in optimal asymmetric cloners the final relation between the two
shrinking factors $\eta_{A}$ and $\eta_{B}$ is in the form of the
equation of an ellipse with an eccentricity depending on
dimension. In the symmetrical case, where
$\eta_A=\eta_B=(d-2)\nu^2+2\nu\sqrt{1+2(d-1)\nu^2}$, a simple
algebra shows that
\begin{eqnarray}
&\nu_{optimal}(d)=\frac{1}{2}\sqrt{\frac{(d^2+4d-4)+(d-2)\sqrt{d^2+4d-4}}{d^3+3d^2-8d+4}},
\end{eqnarray}
from which the fidelity of the optimal phase covariant cloning is
obtained as
\begin{eqnarray}
&F=\frac{1}{d}+\frac{1}{4d}(d-2+\sqrt{d^2+4d-4}),\label{eq19}
\end{eqnarray}
which is in accordance with the result of {\cite{DPCCM}}.

\subsection{Proof of optimality for qubit cloning}
\begin{figure}[tp]
\psfrag{t}[Bc][][0.75][0]{$\theta$} \psfrag{x}[Bc][][0.75][0]{$x$}
\psfrag{y}[Bc][][0.75][0]{$y$} \psfrag{z}[Bc][][0.75][0]{$z$}
\psfrag{r}[Bc][][0.75][0]{$\hat{\mathbf{r}}$}
\includegraphics[width=5.4cm,height=5.2cm]{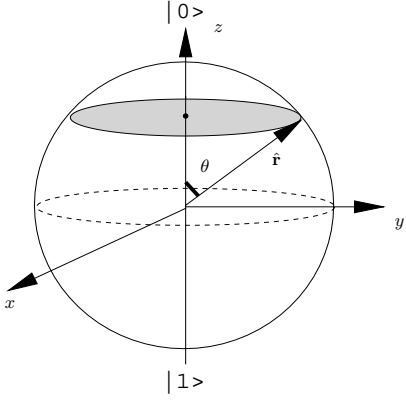}
\caption{The Bloch sphere, and one of its orbits with the polar
angle $\theta$. Any pure state on this orbit is shown by a unit
vector $\hat{\mathbf{r}}$.}
 \label{Bloch}
\end{figure}

In this subsection, we restrict ourselves to the case of $d=2$
(qubit cloning), and study optimality of the cloning
transformations. Optimality of this machine can be proved on the
basis of {\em no-signaling condition}, which has been used
previously in this context {\cite{NS1,NS2,NS3,NS4,NS5}}. In simple
words, this condition states that one cannot exploit quantum
entanglement between two spacelike separated parties for
superluminal communication. However this condition is very
general, a simplified version of that is sufficient for means of
this paper. We show that using this condition (and positive
semidefiniteness of density operators) a trade-off relation
between $\eta_{A}$ and $\eta_{B}$ is found which is saturated by
our cloning transformation.  We devise our proof so that it can
also be used to derive an upper bound on trade-off relations of
cloners of {\em orbital} sates of the Bloch sphere. These states
are simply defined to be the states for which we have $\langle
\psi|\sigma_{z}|\psi\rangle=\cos~\theta$, for a given $\theta$.
Figure~\ref{Bloch} shows an illustration of the Bloch sphere along
with an orbit having the polar angle $\theta$.  The $x$~-~$y$
equator is a special kind of these orbital states with
$\theta=\frac{\pi}{2}$, or equivalently $\langle
\psi|\sigma_{z}|\psi\rangle=0$. Phase-covariant cloners for these
special class of states were initially introduced in
{\cite{Ours}}, and lately were considered for optical
implementation {\cite{Fiurasek}}.

An orbital state, with a given value of $\theta$, can be
represented as
\begin{eqnarray}
&|\psi (\hat{\mathbf{r}})\rangle=\cos
\frac{\theta}{2}|0\rangle+e^{i \phi}\sin \frac{\theta}{2}|1\rangle
~~~(0\leq\phi<2\pi),\label{orbital}
\end{eqnarray}
in which $\hat{\mathbf{r}}=(\sin \theta \cos \phi, \sin \theta
\sin \phi, \cos \theta)$ is a unit vector representing the state
on a given orbit of the Bloch sphere.  Imposing the {\em
universality} condition for the cloning of a given orbit requires
the condition
\begin{eqnarray}
&\rho^{(out)}_{AB}(R\hat{\mathbf{r}})=U(R)\otimes
U(R)\rho^{(out)}_{AB}(\hat{\mathbf{r}})U^{\dag}(R)\otimes
U^{\dag}(R), \hskip 5mm \label{universality}
\end{eqnarray}
for any $\hat{\mathbf{r}}$ on the orbit.  In this relation
{\small{$R\equiv R(\hat{z}, \chi)\in$}}SO(3) is the usual rotation
matrix in 3-dimensional space about $z$-axis through an angle
$\chi$, and {\small{$U(R)=e^{-i\frac{\chi}{2}\sigma_{z}}\in$}}
SU(2) is the corresponding unitary operation on the Bloch vector
(and {\small{ $\sigma_{z}=\left(\begin{smallmatrix}
  1 & 0 \\
  0 & -1 \\
\end{smallmatrix}\right)$}}).  We want to clone (asymmetrically and) universally this
qubit (independent of the Bloch vector $\hat{\mathbf{r}}$ on a
given orbit), in such a way that the reduced density matrices of
the clones ${\small{A(B)}}$,
$\rho^{(out)}_{A(B)}(\hat{\mathbf{r}})$, are of the forms
\begin{eqnarray}
&\rho^{(out)}_{A(B)}=\eta_{A(B)}\rho^{(id)}_{A(B)}+\frac{1-\eta_{A(B)}}{2}1_{A(B)}.
\end{eqnarray}
This equation is usually referred to as the {\em isotropy}
condition. The most general form for the combined output of this
machine, $\rho^{(out)}_{AB}(\hat{\mathbf{r}})$, can be written as
\begin{eqnarray}
&\rho^{(out)}_{AB}(~\hat{\mathbf{r}})=\frac{1}{4}(1\otimes
1+\eta_{A}\hat{\mathbf{r}}.\vec{\sigma}\otimes 1+ 1\otimes
\eta_{B}\hat{\mathbf{r}}.\vec{\sigma}\nonumber\\&\hskip 2mm
+\sum_{j,k=x,y,z}t_{jk}\sigma_{j}\otimes \sigma_{k}),
\label{general}
\end{eqnarray}
where $t_{jk}$'s are real parameters and
{\small{$\vec{\sigma}=(\sigma_{x},\sigma_{y},\sigma_{z})$}}. By
using the identities
\begin{eqnarray}
&U(R)\sigma_{x}U^{\dag}(R)=\cos\chi \sigma_{x}+\sin \chi \sigma_{y}\nonumber\\
&U(R)\sigma_{z}U^{\dag}(R)=\cos\chi \sigma_{y}-\sin\chi \sigma_{x}
\label{identities}
\end{eqnarray}
one gets the following relations for $t'_{jk}$ (the parameters of
$\rho^{(out)}_{AB}(R~\hat{\mathbf{r}})$ )
\begin{eqnarray}
&t'_{xx}=\cos^2\chi t_{xx} +\sin^2\chi t_{yy}-\sin\chi\cos\chi (t_{xy}+t_{yz})\nonumber\\
&t'_{xy}=\cos^2\chi t_{xy} -\sin^2\chi t_{yx}+\sin\chi\cos\chi (t_{xx}-t_{yy})\nonumber\\
&t'_{xz}=\cos\chi t_{xz} -\sin\chi t_{yz}\nonumber\\
&t'_{yx}=\cos^2\chi t_{yx} -\sin^2\chi t_{xy}+\sin\chi\cos\chi (t_{xx}-t_{yy})\nonumber\\
&t'_{yy}=\cos^2\chi t_{yy} +\sin^2\chi t_{xx}+\sin\chi\cos\chi (t_{xy}+t_{yx})\nonumber\\
&t'_{yz}=\cos\chi t_{yz} +\sin\chi t_{xz}\nonumber\\
&t'_{zx}=\cos\chi t_{zx} -\sin\chi t_{zy}\nonumber\\
&t'_{zy}=\cos\chi t_{zy} +\sin\chi t_{zx}\nonumber\\
&t'_{zz}=t_{zz}.
 \label{ts}
\end{eqnarray}
We introduce simpler notations for four special equatorial vectors
below
\begin{eqnarray}
&&|\hat{x}\rangle\equiv (1,0,0)~~~~~ |-\hat{x}\rangle\equiv
(-1,0,0)\nonumber\\ &&|\hat{y}\rangle\equiv (0,1,0)~~~~~
|-\hat{y}\rangle\equiv (0,-1,0). \label{vects}
\end{eqnarray}
No-signaling condition {\cite{NS1,NS2,NS3,NS4,NS5}}, here, reads
as
\begin{eqnarray}
&\rho^{(out)}_{AB}(\hat{x})+\rho^{(out)}_{AB}(-\hat{x})=\rho^{(out)}_{AB}(\hat{y})+\rho^{(out)}_{AB}(-\hat{y}).
\label{no-signaling}
\end{eqnarray}
Putting $\chi=0$ for $|\hat{x}\rangle$ results in
\begin{eqnarray}
&t'_{xx}=t_{xx}, t'_{xy}=t_{xy}, t'_{xz}=t_{xz}, t'_{yx}=t_{yx},
t'_{yy}=t_{yy}, \nonumber\\ &t'_{yz}=t_{yz}, t'_{zx}=t_{zx},
t'_{zy}=t_{zy}, t'_{zz}=t_{zz}. \label{trelation1}
\end{eqnarray}
Similarly inserting the value $\chi=\pi$ for $|-\hat{x}\rangle$
gives
\begin{eqnarray}
&t'_{xx}=t_{xx}, t'_{xy}=t_{xy}, t'_{xz}=-t_{xz}, t'_{yx}=t_{yx},
t'_{yy}=t_{yy}, \nonumber\\ &t'_{yz}=-t_{yz}, t'_{zx}=-t_{zx},
t'_{zy}=-t_{zy}, t'_{zz}=t_{zz},
 \label{trelation2}
\end{eqnarray}
and also $\chi=\frac{\pi}{2}$ for $|\hat{y}\rangle$ results in
\begin{eqnarray}
&t'_{xx}=t_{yy}, t'_{xy}=-t_{yx}, t'_{xz}=-t_{yz},
t'_{yx}=-t_{xy}, t'_{yy}=t_{xx}, \nonumber\\
 &t'_{yz}=t_{xz},
t'_{zx}=-t_{zy}, t'_{zy}=t_{zx}, t'_{zz}=t_{zz},
 \label{trelation3}
\end{eqnarray}
and finally $\chi=\frac{3\pi}{2}$ for $|-\hat{y}\rangle$ gives
\begin{eqnarray}
&t'_{xx}=t_{yy}, t'_{xy}=-t_{yx}, t'_{xz}=t_{yz}, t'_{yx}=-t_{xy},
t'_{yy}=t_{xx}, \nonumber\\ &t'_{yz}=-t_{xz}, t'_{zx}=t_{zy},
t'_{zy}=-t_{zx}, t'_{zz}=t_{zz}. \label{trelation4}
\end{eqnarray}
Considering the above equations and Eq.~(\ref{no-signaling}), the
following conditions are obtained
\begin{eqnarray}
&t_{xx}=t_{yy},~t_{xy}=-t_{yx}. \label{NScondition}
\end{eqnarray}
In this stage, by using the above relations, the explicit form of
the density matrix of the combined outputs can be found as
\begin{eqnarray}
&\rho^{(out)}_{AB}(\hat{x})=\frac{1}{4}(1\otimes
1+\eta_{A}\sigma_{x}\otimes 1+1\otimes
\eta_{B}\sigma_{x}\nonumber\\&+t_{xx}(\sigma_{x}\otimes
\sigma_{x}+\sigma_{y}\otimes \sigma_{y})+t_{xy}(\sigma_{x}\otimes
\sigma_{y}-\sigma_{y}\otimes
\sigma_{x})\nonumber\\&+t_{xz}\sigma_{x}\otimes \sigma_{z}
+t_{yz}\sigma_{y}\otimes \sigma_{z}+t_{zx}\sigma_{z}\otimes
\sigma_{x}\nonumber\\&+t_{zy}\sigma_{z}\otimes
\sigma_{y}+t_{zz}\sigma_{z}\otimes \sigma_{z}). \label{rhox}
\end{eqnarray}
From positive semidefiniteness of this density matrix the
following conditions {\cite{HJ}} are found
\begin{widetext}
\begin{eqnarray}
&-1+\eta_{A}^2+\eta_{B}^2-2\eta_{A}\eta_{B}t_{xx}+2t_{xx}^2+2t_{xy}^2+t_{xz}^2+t_{yz}^2-2t_{xx}t_{xz}t_{zx}+
2t_{xy}t_{yz}t_{zx}+t_{zx}^2\nonumber\\&-2t_{xy}t_{xz}t_{zy}-2t_{xx}t_{yz}t_{zy}+t_{zy}^2+2t_{xx}^2t_{zz}+2t_{xy}^2t_{zz}+t_{zz}^2\leq
0,\label{positive2}\\
& \det(\rho^{(out)}_{AB}(\hat{x}))\geq 0. \label{positive3}
\end{eqnarray}
\end{widetext}
To maximize the value of $\eta_{A}^2+\eta_{B}^2$, it is seen from
Eq.~(\ref{positive2}) that one must take
$t_{xz}=t_{yz}=t_{zx}=t_{zy}=t_{zz}=t_{xy}=0$, which in turn gives
\begin{eqnarray}
&\eta_{A}^2+\eta_{B}^2\leq 1-2 (t_{xx}^2-\eta_{A}\eta_{B}t_{xx}).
\label{firstc}
\end{eqnarray}
From this relation it is concluded that to maximize the value of
the left hand side one should choose
$t_{xx}=\frac{\eta_{A}\eta_{B}}{2}$, hence it follows that
\begin{eqnarray}
&\eta_{A}^2+\eta_{B}^2\leq 1+\frac{\eta_{A}^2\eta_{B}^2}{2}.
\label{firstcc}
\end{eqnarray}
These considerations along with Eq.~(\ref{positive3}), finally
give rise to
\begin{eqnarray}
&(\eta_{A}^2+\eta_{B}^2)^2-4(\eta_{A}^2+\eta_{B}^2)+3\geq 0,
\label{lastc}
\end{eqnarray}
and hence the following upper bound on the qualities of the clones
is obtained
\begin{eqnarray}
 &\eta_{A}^2+\eta_{B}^2\leq1.\label{nosig}
\end{eqnarray}

\begin{figure}[tbp]
\psfrag{x}[Bc][][0.75][0]{$\eta_{A}$}
\psfrag{y}[Bc][][0.75][0]{$\eta_{B}$}
\psfrag{phase}[Bc][][0.75][0]{\hskip 3mm phase cov.}
 \psfrag{univ}[Bc][][0.75][0]{\hskip 3mm universal}
\psfrag{eqq}[Bc][][0.75][0]{\hskip 5mm$\eta_{A}=\eta_{B}$}
\includegraphics[width=7cm,height=5cm]{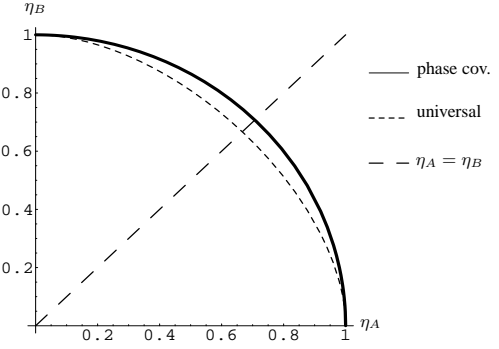}
\caption{Comparison of trade-off ellipses for universal and
phase-covariant cloners in $d$=2. } \label{fig4}
\end{figure}

In the case of qubit cloning, Eqs.~(\ref{eq17a}) and (\ref{eq17b})
reduce to
\begin{subequations}
\begin{eqnarray}
&\eta_{A}(\nu,\xi)=2\nu\sqrt{1-(\nu^2+\xi^2)}\label{eq20a}\\
&\eta_{B}(\nu,\xi)=2\xi\sqrt{1-(\nu^2+\xi^2)}.\label{eq20b}
\end{eqnarray}
\end{subequations}
Obviously we must assume $\nu,\xi\geq 0$ to avoid negative
shrinking factors that are unphysical.  Thus, the only acceptable
value of $\nu$ which optimizes $\eta_{B}$, provided that
$\eta_{A}$=const., is
\begin{eqnarray}
&\nu_{optimal}=\frac{\eta_{A}}{\sqrt{2}}.\label{eq21}
\end{eqnarray}
Therefore, the optimal trade-off relation, now becomes
\begin{eqnarray}
&\eta_{A}^2+\eta_{B}^2=1\label{eq22}
\end{eqnarray}
which is the equation of a unit circle in the $(\eta_A,\eta_B)$
space. This cloner is on the edge of no-signaling condition
(\ref{nosig}).  In corresponding symmetric cloner, for which
$\eta_{A}=\eta_{B}$, Eq.~(\ref{eq22}) gives
\begin{eqnarray}
&F=\frac{1}{2}+\frac{1}{\sqrt{8}},\label{eq23}
\end{eqnarray}
which is the same as the known value {\cite{PCCM}}.
Figure~\ref{fig4} compares the trade-off diagrams for both optimal
universal and optimal phase-covariant cloners.

\subsection{Separability properties of the clones}
In this subsection we want to study entanglement properties of the
outputs of these cloners.  For the symmetric case as has been
shown in {\cite{PCCM}}, using the Peres-Horodecki's positive
partial transposition criterion \cite{Peres,Horod}, only for
optimal phase-covariant cloners two output clones are separable.
We presently want to investigate a similar question in the context
of asymmetric cloning machines. It can be obtained that in $d$=2
partial transposition of the density matrix of the combined
clones, $[\rho_{AB}^{(out)}]^{T_{A}}$, is as follows
\begin{eqnarray}
\hskip -1.5mm &[\rho_{AB}^{(out)}]^{T_{A}} \label{eq24}=\frac{1}{2}\left(%
\begin{array}{cccc}
  \mu^2 & \mu\xi e^{-i\phi} &  \mu\nu e^{i\phi} & 2\nu\xi \\
   \mu\xi e^{i\phi} & \nu^2+\xi^2 & 0 &  \mu\nu e^{i\phi} \\
   \mu\nu e^{-i\phi} & 0 & \nu^2+\xi^2&  \mu\xi e^{-i\phi} \\
  2\nu\xi &  \mu\nu e^{-i\phi} &  \mu\xi e^{i\phi} & \mu^2 \\
\end{array}%
\right)\hskip 2.5mm
\end{eqnarray}
in which we have considered $\alpha_{0}=\frac{1}{\sqrt{2}}$ and
$\alpha_{1}=\frac{e^{i\phi}}{\sqrt{2}}$, and the computational
basis has been used. Eigenvalues of this matrix, in terms of $\nu$
and $\xi$ (after the cancelation of $\mu$ by the normalization
condition), are
\begin{eqnarray}
&\{ \frac{1}{4}(1+2\nu\xi\pm
\sqrt{1+12\nu\xi-16\nu^3\xi+4\nu^2\xi^2-16\nu\xi^3}),\nonumber\\
&\frac{1}{4}(1-2\nu\xi\pm
\sqrt{1-12\nu\xi+16\nu^3\xi+4\nu^2\xi^2+16\nu\xi^3})\}.\hskip
4mm\label{eq25}
\end{eqnarray}
Replacing the optimal values of $\nu=\frac{\eta_{A}}{\sqrt{2}}$
and $\xi=\frac{\eta_{B}}{\sqrt{2}}$, and using Eq.~(\ref{eq22}),
the eigenvalues for the optimal asymmetric phase-covariant cloner
are obtained as
\begin{eqnarray}
&\{0,0,\frac{1}{2}(1-\sqrt{\eta_{A}^2(1-\eta_{A}^2)}),\frac{1}{2}(1+\sqrt{\eta_{A}^2(1-\eta_{A}^2)})\}\label{eq26}
\end{eqnarray}
which, by considering $0\leq \eta_{A}\leq 1$, are all
non-negative. Therefore, for the optimal asymmetric
phase-covariant cloner, like symmetric case, two output copies are
separable{\footnote{Or, equivalently, another good measure of
entanglement named {\em negativity}, defined as: $N(\rho_{AB})=2
~{\text{max}}(0,-\lambda_{\text{min}})$, where
$\lambda_{\text{min}}$ is the minimal eigenvalue of
$\rho_{AB}^{T_{A}}$ {\cite{Vidal}, vanishes.}}. By numerical
analysis of Eq.~(\ref{eq25}) in other cases, we have found that
except in the optimal asymmetric cloners, always at least one of
eigenvalues is negative. Thus it is argued that the special class
of the optimal phase-covariant cloners is unique in that they are
the only cloning machines that give rise to separable output
clones. Therefore, this property of the phase-covariant cloners is
respected in asymmetric case, too.  In the symmetric case, in
which $\eta_{A}=\eta_{B}=\frac{1}{\sqrt{2}}$, Eq.~(\ref{eq26})
gives $\{0,0,\frac{1}{4},\frac{3}{4}\}$ which coincides to the
known value.
\begin{figure}
\psfrag{tau}[Bc][][0.75][0]{$\tau$}
\psfrag{nu}[Bc][][0.75][0]{$\nu$}
\psfrag{theta}[Bc][][0.75][0]{$\theta$}
\includegraphics[width=5.2cm,height=3.5cm]{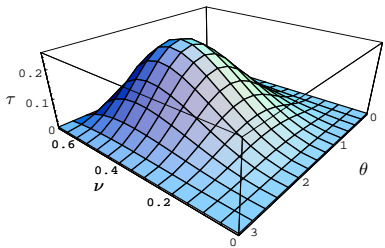}
\caption{Tangle of the total output of the cloning machine vs.
$\nu$ and $\theta$ (in radian). Each $\theta$ is representative of
a specific orbit on the Bloch sphere.}
 \label{tangle3d}
\end{figure}

Moreover, let us study the entanglement properties of the total
output pure state, $|\psi\rangle_{ABX}$.  This state clearly has
three-party entanglement.  Such a similar investigation for the
symmetric cloners has been done in {\cite{Bruss4}}.  Also a
comparison of different cloning machines, in the sense of their
entanglement properties, can be found in {\cite{Masiak}}.\\ There
exists two different inequivalent classes of three-party
entanglement {\cite{Dur}}, namely {\small $W$}- and
\textsc{ghz}-type (shown, respectively, by $|\psi_{W}\rangle$ and
$|\psi_{\textsc{ghz}}\rangle$ ) which their representatives are as
follows
\begin{eqnarray}
&|{\small{W}}\rangle=\frac{1}{\sqrt{3}}(|001\rangle+|010\rangle+|100\rangle),\nonumber\\
&|{\textsc{ghz}}\rangle=\frac{1}{\sqrt{2}}(|000\rangle+|111\rangle).
\label{WG-class}
\end{eqnarray}
For pure states of three qubits, $|\psi\rangle_{ABC}$, there
exists a simple criterion to detect to which class an entangled
state belongs, which is called 3-{\em{tangle}} (or shortly,
tangle) $\tau_{ABC}$ {\cite{Coffman}}.  It can be shown that $\tau
(|\psi_{W}\rangle)=0,$ and $\tau
(|\psi_{{\textsc{ghz}}}\rangle)>0$. A simple calculation shows
that here for the case of qubits
 on an orbit of the Bloch sphere, Eq.~(\ref{orbital}), one
obtains
\begin{eqnarray}
&\tau_{ABX}(\nu, \theta)=4\sin^2\theta~\nu^2(\frac{1}{2}-\nu^2).
\label{tau}
\end{eqnarray}
Figure~\ref{tangle3d} shows a plot of this entanglement measure
vs. $\theta$, and $\nu$.  Also in Fig.~\ref{tangle2d} tangle is
shown for the special case of $x$~-~$y$ equatorial states (which
gain the maximum tangle among all orbits).  As is seen, except the
two special cases $\nu=0$ and $\frac{1}{\sqrt{2}}$, the tangle for
other asymmetric phase-covariant cloners is greater than zero
which indicates that the total outputs of these types of machines
are of \textsc{ghz}-type. The maximum value for the tangle is
attained in the case of $\nu=\frac{1}{2}$ (and subsequently,
$\xi=\frac{1}{2}$) which is the symmetric case.  This property is
reasonable in the sense that in the symmetric case two-party
entanglement of the two clones is zero (clones are separable) and
as well two-party entanglement between the clone {\small $A$} and
the machine {\small $X$} is relatively small
($N(\rho^{(out)}_{AX})_{\text{symm}}\simeq 0.0346$), which implies
that big portion of entanglement is of
three-party{\footnote{Following {\cite{Coffman}}, for pure states
we have $\tau_{AB}+\tau_{AC}+\tau_{ABC}=\tau_{A(BC)}$.  This is a
kind of trade-off between two- and three-party entanglements in
tripartite systems.}} type.
\begin{figure}
\psfrag{tau}[Bc][][0.75][0]{$\tau_{ABX}$}
\psfrag{nu}[Bc][][0.75][0]{$\nu$}
\includegraphics[width=5cm,height=3cm]{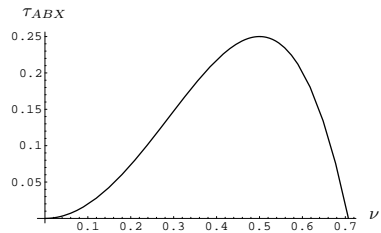}
\caption{Tangle of the total output of the cloning machine vs.
$\nu$. Excluding the cases $\nu=0$ and $\nu=\frac{1}{\sqrt{2}}$
(fully asymmetric cloning), the total output is a
{\textsc{ghz}}-type state. The maximum is attained in symmetric
case, $\nu=\frac{1}{2}$.}
 \label{tangle2d}
\end{figure}

\section{conclusion}{\label{sec4}}
In this paper, we have considered a simple two-parametric family
of optimal asymmetric cloners. In this type of machines if the
quality of a clone (defined by its fidelity) is given then the
quality of the other clone will be the highest possible value.
Thus, the qualities are free to be tuned (up to a complementarity
relation between the two clones). These cloners saturate the
so-called no-cloning inequalities, and therefore are optimal.

 Next, we have
introduced optimal asymmetric cloning transformation for the
special class of phase-covariant (equatorial) states; optimal
asymmetric phase-covariant cloners. An explicit proof, based upon
the no-signaling condition, has been presented for the case of
qubit cloning. The proof can easily be generalized to the
phase-covariant cloners for orbital states (though explicit form
of these cloning transformations has not been mentioned). As well,
for the case of equatorial states, the trade-off between clones
has been obtained explicitly, and entanglement properties of the
clones have been investigated.  It has been argued that among all
inputs, only the equatorial qubits, and only in the case of an
optimal phase-covariant cloner, give rise to separable clones. So,
again as in the optimal symmetric phase-covariant cloners, the
equatorial qubits are unique, and asymmetry in qualities of their
clones respects this special feature. Also three-party
entanglement of the total state of this cloner has been shown to
be of \textsc{ghz}-type, and for symmetric case the value of this
entanglement is maximum.
\begin{acknowledgments}
A. T. R would like to thank V. Karimipour and P. Zanardi for
discussions on quantum cloning, and also acknowledges hospitality
of the I.S.I Foundation (Turin) where some part of the work was
completed.
\end{acknowledgments}
{\em Note added.--} After completion of this work, we became aware
that Lamoureux and Cerf \cite{Lam} have investigated optimal
asymmetric phase-covariant $d$-dimensional cloning by another
method.



\begin{thebibliography}{99}
\bibitem{Dieks} D. Dieks, Phys. Lett. $\mathbf{92A}$, 271 (1982).
\bibitem{WZ} W. K. Wootters and W. H. Zurek, Nature (London) $\mathbf{299}$, 802
(1982).
\bibitem{NoBroadcast}H. Barnum, C. M. Caves, C. A. Fuchs, R. Jozsa, and B. Schumacher, Phys. Rev. Lett. $\mathbf{76}$, 2818
(1996).
\bibitem{Bennett}C. H. Bennett and G. Brassard, in {\em{Proceedings of the IEEE International Conference on Computers,
Systems, and Signal Processing, Bangalore, India}} (IEEE, New
York, 1984), pp. 175-179.
\bibitem{HB1}  V. Bu\v{z}ek and M. Hillery, Phys. Rev. A $\mathbf{54}$, 1844
(1996).
\bibitem{HB2}  V. Bu\v{z}ek and M. Hillery, Phys. Rev. Lett. $\mathbf{81}$, 5003
(1998).
\bibitem{Gisin}  N. Gisin and S. Massar, Phys. Rev. Lett. $\mathbf{79}$, 2153
(1997).
\bibitem{Bruss1} D. Bru\ss, D. P. DiVincenzo, A. Ekert, C. A. Fuchs, C. Macchiavello, and J. A. Smolin, Phys. Rev. A $\mathbf{57}$, 2368
(1998).
\bibitem{Zanardi}P. Zanardi, Phys. Rev. A $\mathbf{58}$, 3484 (1998).
\bibitem{Dlevel} H. Fan, K. Matsumoto, and M. Wadati, Phys. Rev. A
$\mathbf{64}$, 064301 (2001).
\bibitem{Bruss2}D. Bru\ss, A. Ekert, and C. Macchiavello, Phys. Rev. Lett. $\mathbf{81}$, 2598
(1998).
\bibitem{Werner1} R. F. Werner, Phys. Rev. A $\mathbf{58}$, 1827 (1998).
\bibitem{Werner2} M. Keyl and R. F. Werner, J. Math. Phys.
$\mathbf{40}$, 3283 (1999).

\bibitem{Cerf0} N. J. Cerf, Acta Phys. Slov. $\mathbf{48}$,115 (1998).
\bibitem{Cerf1} N. J. Cerf, Phys. Rev. Lett. $\mathbf{84}$, 4497
(2000).
\bibitem{Cerf2} N. J. Cerf, J. Mod. Opt. $\mathbf{47}$, 187 (2000).

\bibitem{HB3}V. Bu\v{z}ek, S. L. Braunstein, M. Hillery, and D. Bru\ss, Phys. Rev. A $\mathbf{56}$, 3446
(1997).
\bibitem{HB4}V. Bu\v{z}ek, M. Hillery, and R. Bednik, Acta Phys. Slov. $\mathbf{48}$,
177 (1998).
\bibitem{Niu}C. -S. Niu and R. B. Griffiths, Phys. Rev. A $\mathbf{58}$, 4377
(1998).

\bibitem{Bruss3}D. Bru\ss, M. Cinchetti, G. M. D'Ariano, and C. Macchiavello, Phys. Rev.
A $\mathbf{62}$, 012302 (2000).
\bibitem{DPCCM}H. Fan, H. Imai, K. Matsumoto, and X. -B. Wang, Phys. Rev. A $\mathbf{67}$, 022317 (2003).
\bibitem{PCCM} H. Fan, K. Matsumoto, X. -B. Wang, and M. Wadati, Phys.
Rev. A $\mathbf{65}$, 012304 (2001).
\bibitem{D'Ariano} G. M. D'Ariano and P. Lo Presti, Phys. Rev. A $\mathbf{64}$, 042308 (2001).

\bibitem{NS1}N. Gisin, Phys. Lett. A $\mathbf{242}$, 1 (1998).

\bibitem{NS2}
  S. Ghosh, G. Kar, and A. Roy, Phys. Lett. A $\mathbf{261}$, 17
  (1999).

  \bibitem{NS3}
  L. Hardy and D. Song, Phys. Lett. A $\mathbf{259}$, 331
  (1999).

  \bibitem{NS4}
  S. Kunkri, Md. M. Ali, G. Narang, and D. Sarkar,
  quant-ph/0201169.

\bibitem{NS5}
  P. Navez and N. J. Cerf, Phys. Rev. A {\textbf
  68}, 032313 (2003).


\bibitem{Ours} V. Karimipour and A. T. Rezakhani, Phys. Rev. A $\mathbf{66}$, 052111 (2002).

\bibitem{Fiurasek} J. Fiur\`{a}\v{s}ek, Phys. Rev. A $\mathbf{67}$, 052314 (2003).

\bibitem{HJ}R. A. Horn and C. R. Johnson, {\em{Matrix Analysis}} (Cambridge University Press, Cambridge, 1985),
Corollary 7.2.4.

\bibitem{Peres}A. Peres, Phys. Rev. Lett. $\mathbf{77}$, 1413
(1996).
\bibitem{Horod} M. Horodecki, P. Horodecki, and R. Horodecki, Phys.
Lett. A $\mathbf{223}$, 1 (1996).

\bibitem{Vidal}K. \.{Z}yczkowski, P. Horodecki, A. Sanpera, and M. Lewenstein, Phys.
Rev. A $\mathbf{58}$, 883 (1998); G. Vidal and R. F. Werner, Phys.
Rev. A $\mathbf{65}$, 032314 (2002).

\bibitem{Bruss4} D. Bru\ss ~and C. Macchiavello, Found. Phys. {\bf 33}, 1617 (2003).
\bibitem{Masiak}P. Masiak, J. Mod. Opt. $\mathbf{50}$, 1873 (2003).

\bibitem{Dur}W. D\"{u}r, G. Vidal, and J. I. Cirac, Phys. Rev. A. $\mathbf{62}$, 062314 (2000).


\bibitem{Coffman} V. Coffman, J. Kundu, and W. K. Wootters, Phys.
Rev. A $\mathbf{61}$, 052306 (2000).

\bibitem{Lam} L. -P. Lamoureux and N. J. Cerf, quant-ph/0410054.



\end{thebibliography}
\end{document}